\documentstyle[aps,prl,epsf]{revtex}
\newcommand{\beq}{\begin{equation}}
\newcommand{\eeq}{\end{equation}}
\newcommand{\beqa}{\begin{eqnarray}}
\newcommand{\eeqa}{\end{eqnarray}}
\newcommand{\evg}{eV_g}
\newcommand{\ed}{\varepsilon_d}
\newcommand{\eox}{\epsilon_{ox}}

\newcommand{\emax}{\varepsilon_{m}}
\newcommand{\est}{\epsilon^{*}}
\newcommand{\eps}{\varepsilon}

\newcommand{\bi}{\bibitem}
\newcommand{\epsbar}{{\bar\eps}}
\newcommand{\trap}{\varepsilon_t}
\newcommand{\calr}{{\cal R}}
\begin{document}
\twocolumn[\hsize\textwidth\columnwidth\hsize\csname@twocolumnfalse%
\endcsname
\title{On the Theory of Metal-Insulator Transitions in Gated
Semiconductors}
\author{Boris L. Altshuler$^{1,2}$ and Dmitrii L. Maslov$^3$}
\address{
$^1$NEC Research Institute, 4 Independence Way, Princeton, NJ 08540\\
$^2$Physics Department, Princeton University, Princeton, NJ 08544\\
$^3$Department of Physics, University of Florida,
P.\ O.\ Box 118440, Gainesville, Florida 32611\\
}
\maketitle
\begin{abstract}
It is shown that recent experiments indicating a metal-insulator
transition in 2D electron systems
can be interpreted in terms of a simple model, in which
the resistivity is controlled by scattering at charged
hole traps located in the oxide layer. The gate voltage changes
the number of charged traps which results in a sharp
change in the resistivity. The observed exponential
temperature dependence of the resistivity in the metallic phase
of the transition follows from the temperature
dependence of the trap occupation number. The model
naturally describes the experimentally observed
scaling properties of the transition and
effects of magnetic and electric fields.
\end{abstract}
\pacs{PACS numbers: 72.10.Bg, 73.20.Dx}
]

Recently, a metal-insulator transition has been observed in low-density
two-dimensional electronic systems -- first in Si MOS
structures \cite{kravchenko,simonian,pudalov,popovic}, and later in other heterostructures \cite{ismail,coleridge,lam,simmons,hanein}.  It
has been found that when the density of 2D electrons $n_{s}$ is below some
critical value $n^{c}_{s}$, cooling causes an increase of the
resistivity $\rho$, while at $n_{s} > n^{c}_{s}$\  the resistivity
{\it decreases} with temperature $T$, i.e., the system
exhibits an unexpected metallic behavior.  The insulating phase
has been found to be rather usual and easy to describe in terms of
variable-range hopping \cite{efros}.  On the contrary, the metallic phase is
anomalous in at least three respects: i)~ the $\rho$(T) - dependence
follows the exponential, i.e., 
$\rho(T) = \rho_{0} + \rho_{1} \exp (-T_{0}/T)$
rather than power-law form;
ii)~$\rho$ drops by about an order of magnitude
when T changes in the range comparable to the  Fermi
energy $\eps_F$ of 2D electrons;  iii)~the metallic state is
quenched by the magnetic field. 

Here, we are not going to discuss existing attempts
\cite{dobroslavljevic,triplet,phillips,belitz,zhang,he,chakravarty,pudalov_model}
to interpret these
experiments. (We found in Refs.~
\cite{dobroslavljevic,triplet,phillips,belitz,zhang,he,chakravarty,pudalov_model} no satisfactory physical explanation of
the substantial drop in the resistivity in a narrow temperature
interval in an obviously nonsuperconducting system).  Instead, we
propose a simple mechanism which seems to naturally explain all
the peculiarities mentioned above.  We believe that our general
idea can be applied to all gated semiconductors.  However, here we
concentrate on Si MOS structures, where the important characteristics
of a 2DEG and of defects are much better known than in other
systems.

A typical $n$-Si MOS structure consists of a metallic gate, 
SiO$_{2}$ layer, and p-type Si substrate.  Strong enough,
positive gate potential attracts electrons which form an
inversion layer at the SiO$_{2}$/Si interface.  It is known \cite{hori}
that due to the oxygen deficit in the oxide, 
there is a substantial concentration of defects close to the
interface, which are capable of
trapping charges.  Even in state-of-the-art devices, there are
more than $10^{12}$ hole traps per cm$^{2}$, such as Si-Si weak
bonds \cite{hori}.  To introduce the idea of our mechanism, we assume all
of the hole traps to be $\star$) characterized by the same energy of
the electron level $\trap$, and $\star \star$)
located at the same distance $z$ from the interface.  We shall
abandon assumption $\star \star$) later on. Effects of a finite width of
the trap band will be discussed elsewhere. 

At T=0, the trap charge (and spin) state is determined by the
chemical potential $\mu$ of the 2DEG.  For $\trap > \mu$, the electron
level is empty, i.e., a hole is trapped.  The trap has a charge
$+e$ and thus {\it causes strong scattering of 2D electrons}.
It is crucial for our theory that the charge state of a trap can
be changed by varying the gate voltage $V_{g}$.  Indeed, the
bigger $V_{g}$ the smaller $\trap = \trap (V_{g})$.
At $V_{g}=V^{*}_{g}(z)$ determined from $\trap(V_{g}^{*}) = \mu$,
the trap captures an electron
(i.e. emits a hole) and is neutralized.
Being neutral
and remote from 2D electrons, the defect cannot scatter them
any more. Neutralization of the oxide charges reduces
resistivity $\rho$ and thus causes an insulator-to-metal
transition.  When T is high $(\gg\vert\trap- \mu\vert)$, roughly half of
the traps are charged.  As a result, $\rho$ is rather high and
depends weakly on both T and $V_{g}$. On the contrary, for $|\mu -
\trap|\leq T$ the density of charged traps behaves as
$\exp[(\trap-\mu)/T]$, resulting in the exponential
$\rho(T)$-dependence [feature (i)].  The transition takes place
for both degenerate and nondegenerate 2DEGs [feature (ii)].
Finally, the magnetic field effect (iii) can be attributed to the
spin freeze-out of holes \cite{smith}: Zeeman splitting favors spin
1/2 (charged) state with respect to the singlet (neutral) state
of the defect.

It should be noted that we neglect here quantum interference of
2D electrons and thus do not attempt to describe the insulating
phase.  However, we will see that even in the classical case
d$\rho/dT$ can change sign due to the $\mu(T)$-dependence. 

Let us now abandon assumption $\star \star$), i.e., take into
account a broad distribution of distances $z$.  In order to
understand why such a distribution does not smear the transition,
we consider the electrostatic energy of an electron in the oxide
$\eps_{e}(z)$.  Given the total oxide thickness $d$ and its
dielectric constant $\eox$, $\eps_e$ can be written as
\begin{equation}
- \eps_{e} (z) = eV_{g}z/d + e^{2}/(2\eox z).
\label{energy}\end{equation}
Here the two terms represent the external electric
field and the image force from the 2DEG, respectively
(charges induced in the gate can be neglected provided that $z\ll d$).
$\eps_{e}(z)$ reaches its maximum
$\emax$ at $z=z_{m}$, where
\begin{equation}
  \emax = -2 \sqrt{eV_{g}\ed};z_{m} = d
  \sqrt{\ed/eV_{g}};\ed \equiv
  e^{2}/(2\eox d).
\label{emax}\end{equation}
$z_{m}$ can also be  expressed through the mean distance $\bar{r}$
between 2D electrons:
\begin{equation}
z_{m} = {\bar{r}}/\sqrt{8} = a_{B} r_{s}/\sqrt{8},
\label{zmax}\end{equation}
where $a_{B} \equiv {\bar{r}}/r_{s}$ is the effective
Bohr radius. Eq.(\ref{zmax}) follows from the relation  between the 2DEG
concentration $n_{s} = 1/(\pi {\bar{r}}^{2})$ and the gate
voltage: $en_{s} = \epsilon_{ox}V_{g}/(4 \pi d)$ \cite{ftn_thr}.
In order to have a
meaning in a macroscopic theory, $z_{m}$ has to exceed the
screening radius of 2DEG (equal to $a_{B}/4$ for Si (001)
surface \cite{ando}).  Therefore in low-density devices $(r_{s} \gg
1/\sqrt{2})$, this length scale is quite legitimate.

Assuming that the double (hole) occupancy of a trap is impossible, the
probability of  a trap to be charged is 
\begin{equation}
  P_{+}(z) = \Big[\frac{1}{2C}\exp\Big(\frac{\mu-\eps_{e}(z)
    -\trap}{T}\Big)
  + 1\Big]^{-1},
\label{prob}\end{equation}
where $C=1$. According to Eqs.(\ref{energy},\ref{prob}), a homogeneous distribution of traps leads
to a distribution of charges which is peaked at $z=z_{m}$, the
width of the peak being
\begin{equation}
\delta z=d\Big[T^{2}\ed(eV_{g})^{-3}\Big]^{1/4} =
z_{m}\Big[T^{2}(\evg\ed)^{-1}\Big]^{1/4}.
\end{equation}
For $d=2000\;$\AA\ \  and $eV_{g}=1\;\rm{eV}$, we get $\ed=1\;{\rm meV}$ and
$\emax=63\;{\rm meV}$, so that $\delta z/z_{m} \simeq
\sqrt{T(K)}/18 \ll 1$, since $T \leq$ 5K. At $T=5\;{\rm K}$,
$z_m\approx 63\;$\AA\ and $\delta z\approx 8\;$\AA.
This sharpness of
the distribution peak in Eq.(\ref{prob}) manifests itself in
a sharp metal-insulator
transition, as $V_{g}$ is varied.

How does a positive charge, separated by a distance
$z\gg a_{B}$ from the 2DEG, affect the resistivity?  It turns out that
a bound localized state is formed with $\xi=z^{3/4}a_{B}^{1/4} <
z$ and $\eps_{b}$ = $- e^{2}/(\est z)$ being the
localization length and the energy of this state, respectively ($\est$
is the effective dielectric constant of the 2DEG).
The trap and bound electron form a dipole, which is oriented perpendicular
to the 2DEG plane.  For $z \sim a_{B} \sqrt{r_s}$ and $\sqrt{r_s}\gg1$,
the (transport) scattering cross-section $\Sigma (\eps,z)$ of
such a dipole for electrons with energy $\eps$ can be evaluated
classically:
\begin{equation}
\Sigma (\eps,z) = 2.74(e^{2}z^{2}/2\epsilon^{*}\eps)^{1/3}.
\label{cross}
\end{equation}
The Drude formula for the resistivity can be written as
\begin{equation}
  \rho=(N_{t}/e^{2}n_{s}) \sqrt{2m^{*} \epsbar} \Sigma (\epsbar,
  z_{m}) \int^{d}_{0} dz P_{+}(z) \Big(\frac {z}{z_{m}}\Big)^{2/3},
\label{drude}\end{equation}
where $N_{t}$ is the total volume concentration of the
traps, $m^{*}$ is the effective mass of the electrons, and
$\epsbar$ has a meaning of their effective energy, 
which can be expressed through the 2DEG Fermi
energy $\eps_F$ via
\beq
\epsbar=\eps_F\Big[\int_0^\infty \frac{d\eps}{4T}\Big(\frac{\eps}
{\eps_F}\Big)^{5/6}\cosh^{-2}\Big(\frac{\eps-\eps_F}{2T}\Big)\Big]^{-6}.
\label{effeps}
\eeq
Eq.~(\ref{effeps}) interpolates between two limits:
$\epsbar\approx T/\Gamma^{6}(11/6)\approx
1.44T$  for
$T\gg\eps_{F}$ , while in the opposite limit
 $\epsbar=\eps_{F}$. For $T=\eps_F$, the effective
energy $\epsbar\approx 2T$. 
\begin{figure}[htb]
\setlength{\unitlength}{1.0in}
\begin{picture}(2.0,5.0)(0.1,0)
\put(0.15,-0.1){\epsfxsize=3.4in\epsfysize=3.0in\epsfbox{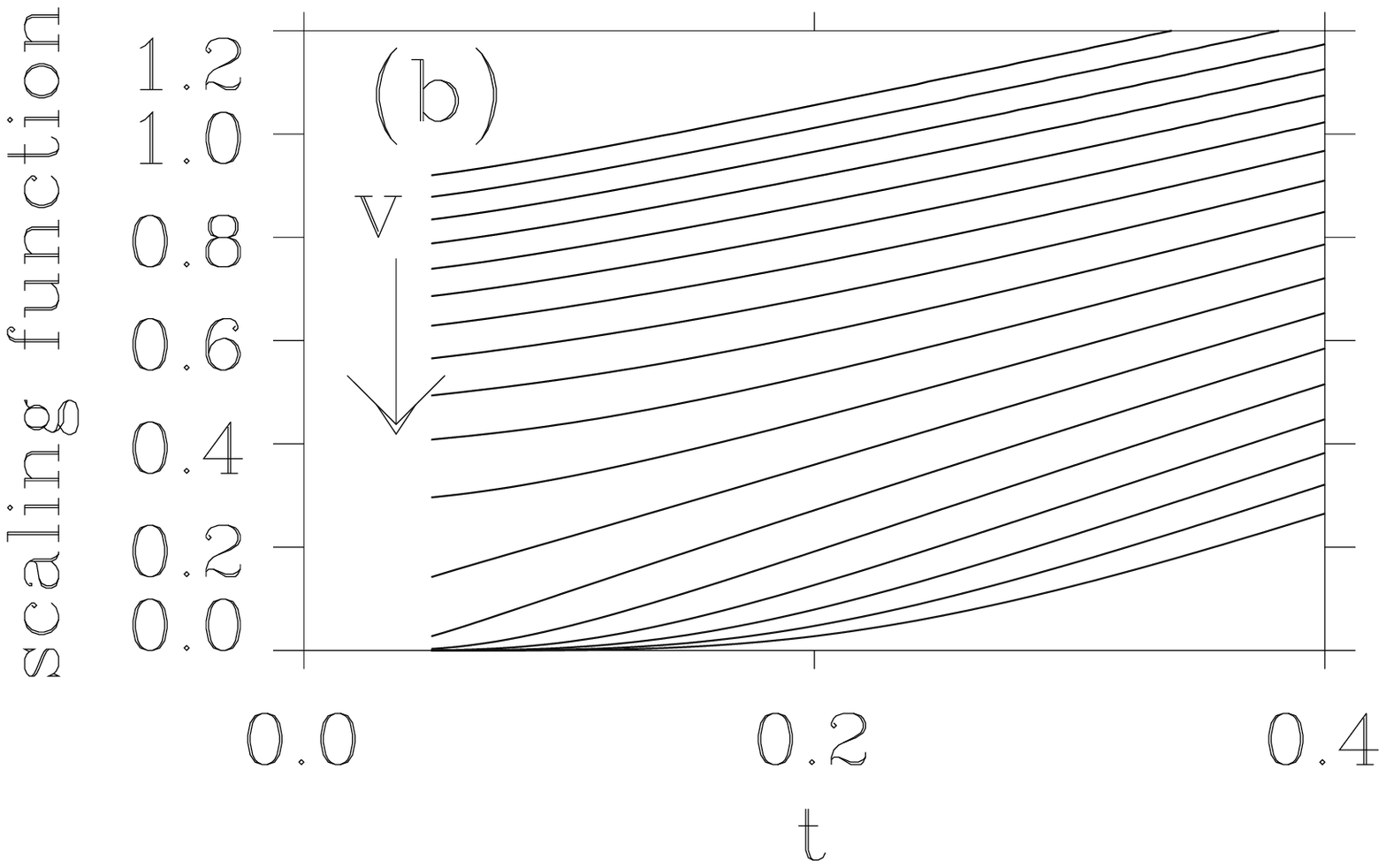}}
\put(0.15,2.2){\epsfxsize=3.4in\epsfysize=3.0in\epsfbox{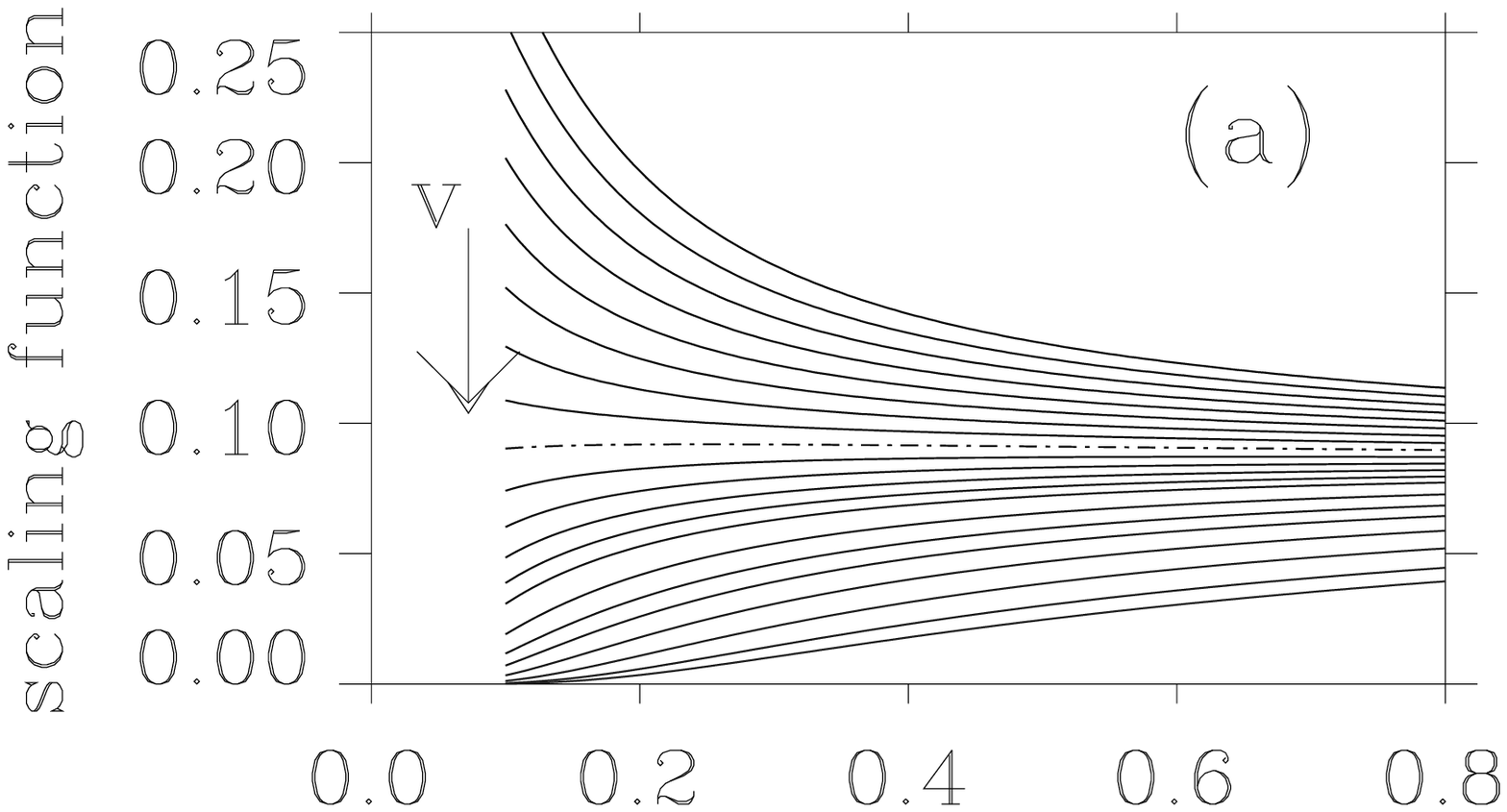}}
\end{picture}
\caption{Scaling function $\calr$ [Eq.~(\ref{rho_c})] 
vs dimensionless temperature $t$  for
several dimensionless gate voltages $v$ [Eq.~(\ref{scale})]. 
$v$ increases in the direction of the arrow. $\eps_F/\ed=0.25$.
{\bf (a)} Case A. $v=-0.2\dots 0.7$. $T_a/\ed=0.04$. Dot-and-dashed
line indicates the transition.
{\bf (b)} Case B.  $v=-1.4\dots 0.7$.}
\label{fig:rho_vs_t}\end{figure}
In the saddle-point approximation, Eq.(\ref{drude}) reduces to
\begin{mathletters}
\beqa
\rho &=& (h/e^{2}) \rho_{0} \calr (V_{g},T); \label{rho_a}\\
\rho_{0} &=& 0.46\sqrt{r_{s}} \big(\est/2 \eox\big)^{1/6}
(N_{t} {\bar{r}}/\pi n_{s}) ({\bar r}/d)^{2/3};\label{rho_b}\\
{\cal R}(V_{g},T) &=&
\Big(\frac{T^{3}\epsbar}{\ed^{4}}\Big)^{1/6}
\int^{\infty}_{0} \frac{dx}{\left[f(T)/2\right] \exp (x^{2} + s) + 1}.\label{rho_c}
\eeqa
\end{mathletters}
\noindent
In Eq.~(\ref{rho_c}), we took into account the $\mu(T)$-dependence:
\begin{equation}
s= \Big[\emax - \trap + \mu(0)\Big]/T;
f(T)=e^{[\mu(T) - \mu(0)]/T}.
\label{var}
\end{equation}
We have to consider two distinct cases: (A) chemical potentials
of the 2DEG and of the Si substrate coincide; (B) the 2DEG is
disconnected from the substrate.  A straightforward
calculation gives
\begin{equation}
f_{A}(T) = (T/T_{a})^{3/4}; \ \ f_{B}(T) = 1-\exp(-\eps_F/T)
\end{equation}
where $T_{a}$ is determined by the acceptor
concentration \cite{smith}. Although case B is more likely
to occur in a real device \cite{pudalov_pc},
we shall concentrate mostly on case A which exhibits
a clear metal-insulator transition even
in a classical model; see below.

\begin{figure}[htb]
\setlength{\unitlength}{1.0in}
\begin{picture}(2.0,2.5)(0.1,0)
\put(0.15,-0.1){\epsfxsize=3.5in\epsfysize=3.0in\epsfbox{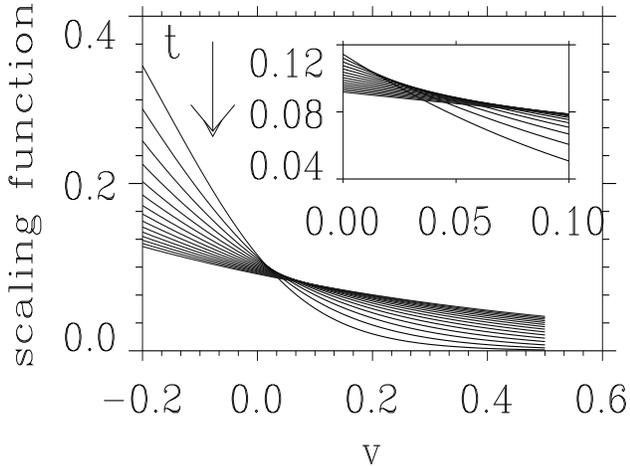}}
\end{picture}
\caption{Case A. Scaling function $\calr$ [Eq.~(\ref{rho_c})]
vs dimensionless gate voltage  $v$  for
dimensionless temperatures $t=0.1\dots 0.6$ [Eq.~(\ref{scale})].
$t$ increases in the direction of the arrow. 
{\it Inset}: a blow-up of the transition region.}
\label{fig:rho_vs_v}\end{figure}
The exponential part of the $\calr(T)$-dependence disappears when $s$,
Eq.~(\ref{var}),
vanishes.  This happens at $V_{g} = V_{g}^{c} = [\trap -
\mu(0)]^2/4e\ed$ and thus $V_{g}^{c}$ defines the
transition point. The asymptotic behavior of $\calr$ away
from this point is
\beqa 
\calr_{A,B}=\left(\frac{T^{3}\epsbar}{\ed^{4}}\right)^{1/6}\!\!\times
\left\{
\begin{array}{ll}
\sqrt{\pi}\Omega,\; {\rm for}\;\Omega\ll 1;\label{asympt}\\
\ln^{1/2}\Omega\;, {\rm for}\;\Omega\gg 1.\\
\end{array}
\right.
\eeqa
where $\Omega\equiv[2/f(T)]e^{-s}$.
The distance from the transition can be measured by $\delta \equiv
(V_{g} - V_{g}^{c})/V_{g}^{c}$.  Provided $\delta^{2} \ll
4T^{2}/(\ed \evg)$, variable $s$ in Eq.(\ref{var}) acquires a
scaling form
\begin{equation}
s \approx \sqrt{\ed \evg^c} (\delta/T) \equiv v/t;  t \equiv
T/\ed;v \equiv \delta \sqrt{\evg^c/\ed}.
\label{scale}\end{equation} 
The $\calr(T)$-dependence in the scaling region is
shown in Fig.~\ref{fig:rho_vs_t}. 
For $v\gg t$, the system is in the
\lq\lq metallic\rq\rq\/ phase characterized
by $\calr_{A,B}$ exponentially decreasing with $t$.
Due to the $\mu(T)$-dependence, 
$d\calr_A/dt$ changes sign at some $v$ slightly bigger
than zero, exhibiting
thus a metal-to-insulator transition. For larger
negative $v$ [not shown in Fig.~\ref{fig:rho_vs_t}(a)],
the $\calr_A(t)$-dependence saturates.
At $\evg^c = 1\;\rm{eV}$, $\ed=1\;{\rm meV}$, and $T =5\;\rm{K}$, we predict
critical behavior for $\vert \delta \vert \leq 0.01$.  This is
consistent with experiments \cite{kravchenko,simonian,pudalov,popovic}.

In case B, there are two distinct regions: exponentially
decreasing and $t$-independent $\calr_B$, the crossover
between the two occuring for $|v|\simeq t$.
Quantum interference effects should result in localization,
converting $T$-independent $\rho$
into an exponentially diverging one; the
metal-insulator transition in this case
will be discussed elsewhere.

At the transition, $\calr_A\approx 0.1$
and $\calr_B\approx 1$(cf. Fig.~\ref{fig:rho_vs_t}).
At the same time, $n_{s} \simeq
10^{11}\;{\rm cm}^{-2}$ and ${\bar r}\simeq 100\;$\AA\ \ for
$d=2 \cdot 10^{-5}\;{\rm cm}$ and $eV_g=1\;\rm{eV}$.
Estimating $N_{t}
{\bar{r}}\simeq 10^{12}\;{\rm cm}^{-2}$ and $r_{s} \simeq 10$, we obtain
for the resistivities at the transition $\rho^c_{A}\simeq 0.1 h/e^2$
 and $\rho^c_{B}\simeq h/e^2$.
These values are within the experimentally observed range \cite{pudalov}b.

As Fig.~\ref{fig:rho_vs_v} shows, the transition
between the insulating  and  metallic phases in case A
is very well-defined, despite the fact that ${\cal R}$ does
not solely depend on the scaling variable $v/t$.
Closer inspection of the transition
region (Fig.~\ref{fig:rho_vs_v}, inset)
reveals however that the transition occurs over a
finite range of $v$ rather than at a single point.
\begin{figure}[htb]
\setlength{\unitlength}{1.0in}
\begin{picture}(2.0,2.5)(0.1,0)
\put(0.15,-0.1){\epsfxsize=3.0in\epsfysize=3.0in\epsfbox{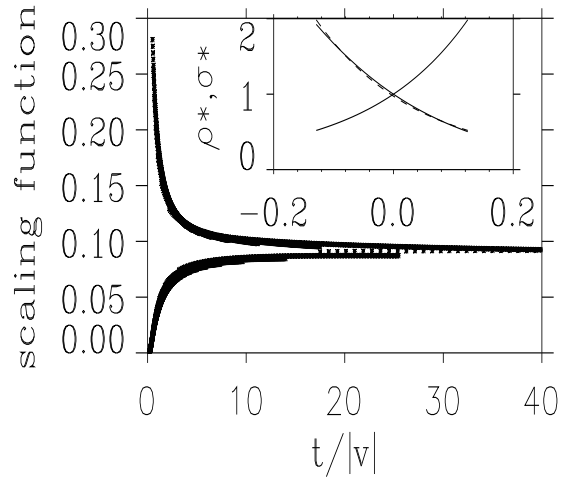}}
\end{picture}
  \caption{Case A. Data collapse in $\calr$ plotted vs $t/|v|$.
{\it Inset}: \lq\lq Duality\rq\rq\/ between 
\lq\lq resistivity\rq\rq\/ $\rho^*=\calr_A/\calr^c_A$
and $\sigma^*=1/\rho^*$ plotted as a function
of $\delta$ [Eq.~(\ref{scale})]. Solid downward: $\rho^*(\delta)$.
Solid upward: $\sigma^*(\delta)$. Dashed: $\sigma^*(-\delta)$.}
\label{fig:scaling}\end{figure}
Figure \ref{fig:scaling} depicts the (approximate) data collapse for $\calr$ plotted as
a function of $t/|v| = T/T_{0}$ with $T_{0} = \vert\delta\vert
\sqrt{\ed\evg^{c}}$.
The inset of Fig.~\ref{fig:scaling} demonstrates
the \lq\lq duality\rq\rq\/ feature, i.e., the symmetry 
between the
resistivity $\rho$ in the insulating phase and the conductivity $\sigma$ 
in the metallic one. Experimentally, a similar collapse was
achieved  for $T_{0} \propto \vert\delta\vert^{a}$ [in the quantum phase
transition theory (QPTT), $a=\nu z$].  In all of the experiments, except
Ref.~\cite{simmons}, $a$ is close to $1$, i.e. to our prediction.

An additional insight comes from nonohmic measurements.  In
Ref.~\cite{kravchenko}c, the dependence of $\rho$ on the source-drain voltage
$V_{SD}={\cal E} L$ (where ${\cal E}$ is the electric field, and $L$
is the source-drain distance) was also found to be a scaling one
$\rho = \rho({\cal E}/{\cal E}_{0}$) with ${\cal E}_0 \propto
\vert\delta\vert^{b}$.  We believe that this ${\cal E}$-dependence can be
attributed to simple heating.  Indeed, the effective temperature of
electrons $T^{*}$ is determined by the energy balance.  For 
strong enough electric field, i.e, when $T^{*}\gg T$, and for
2D electrons,
\begin{equation}
e{\cal E}\sqrt{D(T^{*}) \tau_{eph}(T^{*})} =(\pi/\sqrt{6}) T^{*}.
\label{balance}\end{equation}
Here $D(T)$ is the diffusion constant of electrons at
temperature $T$, and $\tau_{eph}(T) \propto T^{-p}$ is
the relaxation time of the electron temperature, which
we assume to be determined by electron-phonon scattering \cite{ftn_heat}. One can
check that if $\rho(T)$ and $D(T)$ obey a scaling law $1/D(T)
\propto \rho(T) = F(T/T_{0})$, the ${\cal E}$-dependence of the
resistivity is also a scaling-like:
$\rho$ = $G({\cal E}/{\cal E}_{0})$, where ${\cal E}_{0} =
T_{0}^{\alpha} \propto \vert\delta\vert^{\alpha a}$ with $\alpha = 1 + p/2$,
and where function $G$ is obtained by solving Eq.~(\ref{balance})
for given $F$.
If $p=3$ (as it is the case for good metals), $\alpha=2.5$.
Experimental value of $\alpha=b/a$ \cite{kravchenko} is $\simeq 2.25$.
This discrepancy can easily be explained by $p$ being smaller than
$3$.  On the other hand, QPTT predicts $\alpha = 1+z^{-1}$, i.e.
$\alpha$=2 at $z=1$.

One can check that the strong heating regime
is realized under the conditions of Ref.~\cite{kravchenko}c, if
\beq
\tau_{eph}>0.1{\rm ms}/\left[{\cal E}({\rm mV/cm})\right]^2.
\eeq
Strong heating
of a 2DEG has recently been observed 
in a Si MOSFET device \cite{prober} similar in its parameters
to that used in Refs.~\cite{kravchenko,simonian,pudalov,popovic}
\begin{figure}[htb]
\setlength{\unitlength}{1.0in}
\begin{picture}(2.5,2.5)(0.1,0)
\put(0.15,-0.1){\epsfxsize=3.2in\epsfysize=3.0in\epsfbox{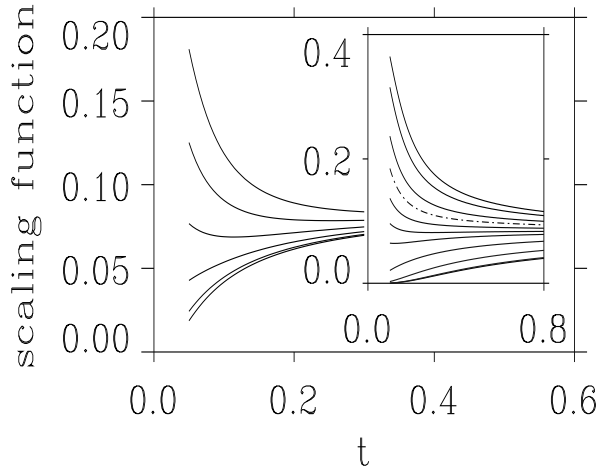}}
\end{picture}
\caption{Case A. Scaling function $\calr$ vs
temperature $t$ for Zeeman splittings
$E_Z=(0, 0.04, 0.08, 0.12, 0.16, 0.2)\times \ed$ ($E_Z$ increases from the bottom to the top curves).
$v=0.1$.
{\it Inset}: The same as in Fig.~\ref{fig:rho_vs_t}a 
but for $E_Z=0.15\ed$.} 
\label{fig:zeeman}\end{figure}

 We now turn to the effect of a magnetic field.
Consider a hole trap, e.g., a Si-Si weak bond
\cite {hori}. Such a trap can find itself in one of the three states
with energies $E_i$; $i=1,2,3$.
For $i=1$ two electrons occupy the bond. This is supposed to be a neutral
($Q=0$), singlet  ($S=0$) state. State 2 (3) has one spin down (up) 
electron
on the bond. Accordingly, $Q=+1$ and $S=1/2$ for both states 2 and 3.
A magnetic field splits the doublet: 
$E_1  - E _{2(3)} = \trap\pm E_Z$, where $E_Z$ 
is the Zeeman splitting. As a result, at  given $T$ and $V_g$
probability $P_+$ to find a trap in a $Q=+1$ state increases with $E_Z$,
i.e., with the magnetic field: one  should substitute 
$C=\cosh(E_Z /T)$  instead of 1 into  Eq.(4).
This results in a magnetoresistivity demonstrated in Fig.~\ref{fig:zeeman}.

We benefited greatly from discussions with 
M.\ E.\ Gershenson and V.\ M.\ Pudalov. We would also like
to thank I.\ L.\ Aleiner, L.\ I.\ Glazman, J.\ Graybeal,
S.\ M.\ Kravchenko, K.\ Muttalib, M.\ Reizer,
and O.\ A.\ Starykh for their valuable comments.
The work of D.\ L.\ M. was supported by NSF  DMR-970338.

\end{document}